\documentclass[structabstract]{aa}

\usepackage{graphicx}
\usepackage{aalongtable}
\usepackage{lscape}
\usepackage{natbib}
\newcommand{\gr}{{$\gamma$-ray}}
\newcommand{\lsim}{{\lower.5ex\hbox{$\; \buildrel < \over \sim \;$}}}
\newcommand{\gsim}{{\lower.5ex\hbox{$\; \buildrel > \over \sim \;$}}}
\newcommand{\nupeak}{$\nu_{\rm peak}$}
\newcommand{\nufnupeak}{$\nu_{\rm peak}$F$_{\nu_{\rm peak}}$}

\begin{document}
   \title{1WHSP: an IR-based sample of  $\sim$1,000 VHE $\gamma$-ray blazar candidates }

   \author{
                     B. Arsioli  \inst{1,2,4} 
		  \and			  
		  B. Fraga    \inst{1,2,4}		  
		  \and
		  P. Giommi   \inst{3,4}	
		   \and
		  P. Padovani \inst{5,6}
		   \and
		  P.M. Marrese  \inst{3}            
          }
        
        \institute{Sapienza Universit\`a di Roma, ICRA, Dipartimento di Fisica, Piazzale Aldo Moro 5, I-00185 Roma, Italy 
        \and
        Universit\'e de Nice Sophia Antipolis, Nice, CEDEX 2, Grand Chateau Parc Valrose,    \\        
        \email{bruno.arsioli@asdc.asi.it}
	\email{bernardo.machado@icra.it}             
        \and 
        ASI Science Data Center, ASDC, Agenzia Spaziale Italiana, via del Politecnico snc, 00133 Roma, Italy
	    \and
	    ICRANet-Rio, CBPF, Rua Dr. Xavier Sigaud 150, 22290-180 Rio de Janeiro, Brazil
       \and
	    European Southern Observatory, Karl-Schwarzschild-Str. 2, D-85748 Garching bei M\"unchen, Germany
	    \and
	    Associated to INAF - Osservatorio Astronomico di Roma, via Frascati 33, I-00040 Monteporzio Catone, Italy
	          }

\date{Accepted: April 10, 2015}

\abstract
{Blazars are the dominant type of extragalactic sources at microwave and at $\gamma$-ray energies. In the most energetic part of the electromagnetic spectrum (E $\gsim 100$ GeV) a large fraction of high Galactic latitude sources are blazars of the  High Synchrotron Peaked (HSP) type, that is BL Lac objects with synchrotron power peaking in the UV or in the X-ray band.  
Building new large samples of HSP blazars is key to understand the properties of jets under extreme conditions, and to study the demographics and the peculiar cosmological evolution of these sources.
}
{
HSP blazars are remarkably rare, with only a few hundreds of them expected to be above the sensitivity limits of currently available surveys, some of which include hundreds of millions of sources. To find these very uncommon objects, we have devised a method that combines ALLWISE survey data with multi-frequency selection criteria.} 
{The sample was defined starting from a primary list of infrared colour-colour selected sources from the ALLWISE all sky survey database, and applying further restrictions on  IR-radio and IR-X-ray flux ratios. Using a polynomial fit to the multi-frequency data (radio to X-ray) we estimated synchrotron peak frequencies and fluxes of each object.} 
{We assembled a sample including 992 sources, which is currently the largest existing list of confirmed and candidates HSP blazars. All objects are expected to radiate up to the highest $\gamma$-ray photon energies. In fact, 299 of these are confirmed emitters of GeV  \gr\ photons (based on Fermi-LAT catalogues), and  36 have already been detected in the TeV band. The majority of sources in the sample are within reach of the upcoming Cherenkov Telescope Array (CTA), and many may be detectable even by the current generation of Cherenkov telescopes during flaring episodes.
The sample includes 425 previously known blazars, 151 new identifications, and 416 HSP candidates (mostly faint sources) for which no optical spectra is available yet. The full 1WHSP catalogue is on-line at http://www.asdc.asi.it/1whsp/ providing a direct link to the SED building tool where multifrequency data for each source can be easily visualised.
}  
{}


 
 \maketitle
%

\section{Introduction}

Blazars are a class of active galactic nuclei (AGN) characterised by rapid and large amplitude 
spectral variability, assumed to be  due to the presence of a relativistic jet pointing very close to the 
line of sight. The emission of these objects is  non-thermal over most or the entire electromagnetic spectrum, from radio frequencies to hard $\gamma$-rays. Usually the observed radiation shows extreme properties, mostly coming from relativistic amplification effects. The observed Spectral Energy Distribution (SED) presents a general shape composed of two bumps, one in the infrared (IR) to soft X-ray band and the other one in the hard X-ray to $\gamma$-rays. According to the standard picture \citep[e.g.][]{giommisimplified}, the first peak is associated with the emission of synchrotron radiation due to relativistic electrons moving in a magnetic field, and the second peak is mainly associated with synchrotron photons that are Inverse-Compton (IC) scattered to higher energies by the same relativistic electron population that generates them (Synchrotron Self Compton model, SSC). The seed photons undergoing IC scattering can also come from outside regions, like the accretion disk and the broad line region, and can add an extra ingredient (External Compton models, EC) for modeling the observed SED. 

If the peak frequency of the synchrotron bump ($\nu_{\rm peak}$) in $\nu$ - $\nu$F$_{\nu}$ space is 
larger than $10^{15}~$Hz, a blazar is usually called High Synchrotron Peaked (HSP) BL Lac, or HBL in the original BL Lac classification of \cite{padgio95}, which was later extended to all blazars 
 by \cite{abdo10}. HSPs are considered \textit{extreme} sources since the Lorentz Factor of the electrons radiating at the peak of the synchrotron SED ($\gamma_{peak}$) are the highest ones within the blazar population. Considering a simple SSC model where $\nu_{peak}=3.2 \times 10^6  \gamma^{2}_{peak} B \delta $ ~ \citep{giommisimplified}, assuming $B=0.1$ Gauss and Doppler factor $\langle \delta \rangle =10$, an HSP characterised by $\nu_{peak} = 10^{15}-10^{17}$Hz demands $\gamma_{peak} \approx 10^4-10^5$. In addition, observations have shown that HSPs are also bright and extremely variable sources of high energy  
TeV photons\footnote{http://tevcat.uchicago.edu} and that they may be the dominant component of a putative extragalactic TeV background \citep{giommicmb}.

The very high energy (VHE) $\gamma$-rays from blazars may be absorbed due to interaction with 
extragalactic background light (EBL) photons  ($\gamma_{VHE} + \gamma_{EBL} \rightarrow e^+ + 
e^- $). The resulting electron positron pairs cool by scattering cosmic microwave background 
(CMB) photons to \gr\ energies, which are offset by a small angle w.r.t. the line of sight when the pairs are deflected in the possible presence of intergalactic magnetic fields (IMFs) \citep[e.g.,][]{dermer}. Studying the development of the cascade through intergalactic distances may provide a tool to constrain the EBL fluxes at the IR range \citep{HESS_EBL} and also imposes lower limits to the IMF. The attenuation due to the EBL may leave a characteristic imprint which is dependent on the redshift of the source and the observed energy band, but a true understanding of such a process demands a clear description of the intrinsic SED generated by the AGN's central engine. 

There is a clear need for a large number of TeV targets in order to gain insight into the underling 
physics. It is therefore important to build a large sample of HSP objects to provide bright 
targets for $\gamma$-ray and TeV detections. This will also permit the study of variability in 
different energy bands to search for fundamental correlations. Within the motivations for identifying 
extreme AGNs there is also the possibility of studying jet properties in extreme conditions and 
determining the population distribution of HSPs. Since AGNs can be detected in a broad range of 
redshifts, extreme bright blazars may also be an efficient tool for studying cosmological structures 
formation and evolution \citep{puccettixmm}.

The SEDs of HSPs are so extreme that no other type of extragalactic sources exhibit similar features. Imposing selection rules (like colour-colour selection and multi-wavelength 
flux ratio limits) that are consistent only with the SED of HSPs, allow us to identify these sources amongst the much larger number of different objects coming from all-sky surveys, and therefore build representative samples with high 
selection efficiency. 
        
Following this guiding idea, the present paper is organised as follows: Section \ref{sample} describes in detail our selection procedure and its efficiency; Section \ref{catalog} presents the HSP sample, its associations with other catalogues, and the derivation of lower limits on the redshifts of some sources; Section \ref{discussion} discusses some characteristics of our HSP sample like the redshift and \nupeak~distribution and the IR LogN-LogS. Also, by placing our sources in a log(L$_{bol}$) vs. log($\nu_{\rm peak}$) plot, we discuss the blazar sequence scenario. A study of the detected/undetected population of TeV sources is also performed and we discuss the likelihood of TeV detectability of our sources. 

Throughout this paper we adopt a Flat-LCDM cosmology with the following parameters: $\Omega_M$=0.315, $\Omega_\Lambda$=0.685, 
H$_0$ = 67.3 km s$^{-1}$ Mpc$^{-1}$ \citep{Planck}. 

\section{Building a large sample of HSP blazars}\label{sample}
\label{SampleDefinition}

An effective way of building large samples of blazars is to work with multi-frequency data, specially from all-sky surveys, and apply selection criteria based on spectral features that are known to be specific of blazar SEDs \citep[e.g.][]{SedentaryI,dxrbs}. Studies performed by \cite{massaro_color} and \cite{dabruscocolors} based on data from the Wide-field Infrared Survey Explorer \citep[WISE,][]{WISE}, have shown that blazars tend to concentrate in a distinct region of the IR colour-colour diagram, named by the authors \textit{the WISE blazar strip}. 
To this purpose we used the latest version of the WISE catalogue (ALLWISE), including 747 million objects, and
the complete sample of about 150 HSPs from the \textit{Sedentary Survey} \citep{SedentaryI,SedentaryIII} which represents well, both in terms of overall SED shape and fluxes, the type of  extreme blazars that we want to select in this paper. The 
\textit{Sedentary Survey} selects high Galactic latitude 
($\vert b \vert  >20^{\circ}$) sources characterised by a very large X-ray to radio flux ratio ($f_{x}/f_{r} \geq 3 \times 10^{-10}erg~cm^{-2} s^{-1} Jy^{-1}$) by means of a multi-frequency method. It is radio flux density limited  and complete above a flux density of $f_{r} \geq ~3.5 mJy$ at 1.4GHz. In addition, we have also considered the IR colours of the HSPs listed in the presently largest compilation of certified blazars: the BZCAT catalogue \citep{5BZcat}. To avoid problems with source confusion and galactic absorption we concentrated in areas of the sky at $\vert b \vert >20^{\circ}$.

\begin{figure}[h]
   \centering
      \includegraphics[width=1.0\linewidth]{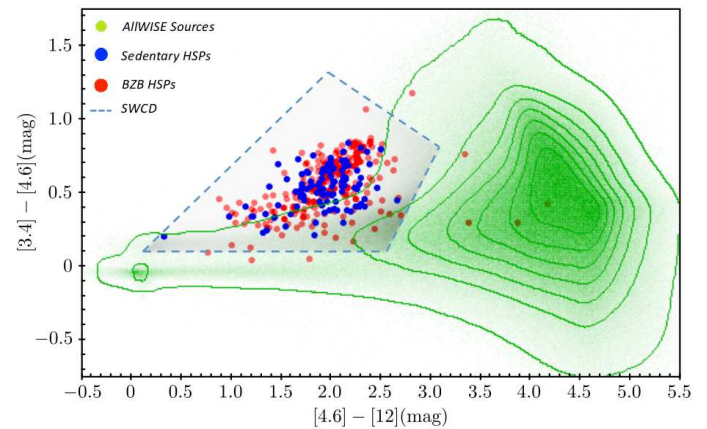}
     \caption{The ALLWISE colour-colour diagram (equivalent to Figs. 1 and 2 of  \cite{massaro_color}) with superposed HSP blazars from the Sedentary survey (blue points) and from BZCAT (red points). The \textit{Sedentary WISE Colour Domain} (SWCD) is delimited by the dashed lines. For both Sedentary and BZCAT sources, as well for the sources in the SWCD, we only consider objects which meet the requirement W$_{3 ~ snr}\geq~$2.0.}
      \label{SWCD}
\end{figure}

Figure \ref{SWCD} shows the peculiar IR colours of HSPs in both the Sedentary (blue points) and the BZCAT samples (red points)  compared to the bulk of WISE IR sources (green points and contours). The special position of blazars reflects the IR spectral slope of the non-thermal radiation from the jet, typically a power law with energy index in the range $0.4-0.8$. However, in several blazars, like e.g. MKN 421 and MKN 501 as well as in many lower luminosity objects, the non-thermal IR continuum is contaminated by the presence of the host galaxy (normally a giant elliptical).

\begin{figure}[h]
 \centering
  \hspace*{-0.2cm}\includegraphics[width=1.0\linewidth]{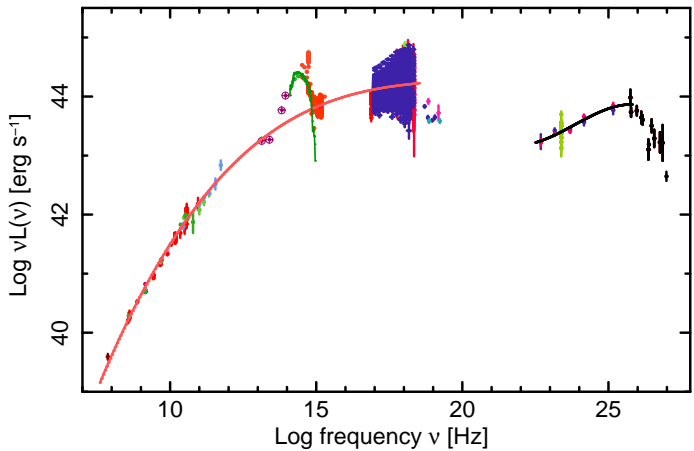}
   \caption{The SED of MKN 501 (1WHSP J165352.2+394536) at z=0.033, a TeV detected HSP showing strong variability in X-rays (Swift-XRT, indigo dots). The giant elliptical galaxy template is represented as green solid-line, at z=0.033. The cross-circle marker represents the four IR WISE channels, and the contamination from W1 (3.4$\mu m$) and W2 (4.6$\mu m$) is clearly seen. The solid lines represent the average non-thermal emission.
}
    \label{GalaxyContribution}
\end{figure}

This is illustrated in Fig. \ref{GalaxyContribution}, where the giant elliptical galaxy template, appearing as a green line, is clearly dominant in the IR to UV range, extending the distribution of HSPs
to the lower left corner of the IR colour-colour plane. In such cases, the thermal component was subtracted before fitting the SED (see below). The plot also shows how the very large variability observed in many objects influences the determination of  \nupeak\ and \nufnupeak.

Therefore considering the existence of such cases, the sources may populate a wider region in the colour-colour diagram, larger than expected for a population composed only by non-thermal-dominated IR emitters. Figure \ref{SWCD} shows that HSP blazars, as represented by the Sedentary Survey and the BZCAT HSPs, still populate a reasonably small and isolated area of the ALLWISE colour-colour plane. 

We have then identified the part of the WISE IR colour-colour space that includes all the Sedentary sources, and nearly all the HSPs  listed in BZCAT, that meet the requirement of being detected by WISE in the 3.4$\mu$m (W$_{1}$), 4.6$\mu$m (W$_{2}$), and 12$\mu$m (W$_{3}$) channels with signal-to-noise ratio (snr) $\geq 2.0$. This region corresponds to the area delimited by the dashed lines in Fig. \ref{SWCD}, which is defined by the following corners:
\\ 

\noindent Corner 1 ($c_{4.6-12\mu m}$ = 0.119, $c_{3.4-4.6\mu m}$ = 0.100), \\ 
Corner 2 ($c_{4.6-12\mu m}$ = 2.552,  $c_{3.4-4.6\mu m}$ = 0.100), \\
Corner 3 ($c_{4.6-12\mu m}$ = 3.090,   $c_{3.4-4.6\mu m}$ = 0.800), \\ 
Corner 4 ($c_{4.6-12\mu m}$ = 2.000, $c_{3.4-4.6\mu m}$ =  1.300), \\ 

\noindent where $c_{3.4-4.6\mu m}=m_{[3.4\mu m]}-m_{[4.6\mu m]}$ \\
 and $c_{4.6-12\mu m}=m_{[4.6\mu m]}-m_{[12.0\mu m]}$\footnote{WISE magnitudes are in the Vega system.}
\\
\\

This area is a compromise between the desire of selecting as many as possible HSPs,  taking into account that in some cases the host galaxy might contaminate the IR colors, and the need to keep the number of IR candidates to a manageable level. With this in mind  we have explored the upper left corner of Fig. \ref{SWCD}, where there are only a few known HSPs but the density of sources is low compared to other quadrants. 
From now on we will call this region of the WISE colour-colour plane the \textit{Sedentary Wise Colour Domain} (SWCD). 
About 36 sources in the Sedentary survey do not meet the requirement of being detected with signal-to-noise ratio (snr) $\geq 2.0$ in all the WISE channels considered, and these cases with unreliable IR colours due to bad photometry are not shown in Fig. \ref{SWCD}. Clearly this leads to some incompleteness in general, and particularly at faint IR fluxes, where the strong requirement of detection in three bands is frequently not met. 

The SWCD includes over 4.8 million objects that are above the Galactic plane ($\vert b \vert  >20^{\circ}$, so that extinction at IR frequencies is negligible) and detected with snr $\geq 2.0$ in all the 3.4 $\mu m$, 4.6 $\mu m$, and the 12 $\mu m$ WISE channels. Although the size of this initial sample of IR colour selected candidates is only about 1.1\% of that of all the ALLWISE sources located at $\vert b \vert  >20^{\circ}$, it 
still includes a very large fraction of non-blazar sources, and it is far too large to be considered for optical spectroscopy follow up. 

To remove as many as possible non-blazar objects from this initial set of ALLWISE candidates we have imposed a number of additional restrictions based on the well-known broad-band spectral peculiarities of blazars. This was done by performing a cross-match between the position of the WISE colour selected objects with a number of radio \citep[NVSS, FIRST and SUMSS:][]{condon_nvss,becker_first,mauch_sumss} and X-ray \citep[IPC, ROSAT BSC and FSC, XMM, SWIFT:][]{IPC,ROSATBSC,ROSATFSC,XMM,SWIFT,deliaswift} catalogues and then applying the following constraints:
\\

~0.05 $<  \alpha_{1.4GHz-3.4\mu m}   <$ 0.45   (1)\\
\indent ~0.4~~ $<  \alpha_{4.6\mu m-1keV}  ~~~ < $ 1.1  ~~(2)  \\
\indent -1.0~~ $< \alpha_{3.4 \mu m-12.0\mu m}  <$ 0.7  ~~(3)  

where $\alpha_{\nu1- \nu2}=-\frac{\log(f_{\nu1}/f_{\nu2})}{\log(\nu_1/\nu_2)}$. \\

A radius of 0.1 arcmin was adopted for the cross-correlations unless the positional uncertainty of a catalogued source (as e.g. in the case of many X-ray detections in the RASS survey) was 
reported to be larger than 0.1 arcmin. In such cases we used the 95\% uncertainty radius (or ellipse major axis)  of each source as maximum distance for the cross-match. 
In addition, to avoid selecting objects with misaligned jets (which are expected to be radio-extended), the spatial extension of radio counterparts (as reported in the original catalogues) was limited to 1 arcmin. 
This procedure was carried out whenever possible, based on the 1.4 GHz radio image from NVSS, which includes the entire sky north of $-40^{\circ}$ declination.

\begin{figure}[ht]
 \centering
   \hspace*{-0.0cm}\includegraphics[width=1.0\linewidth]{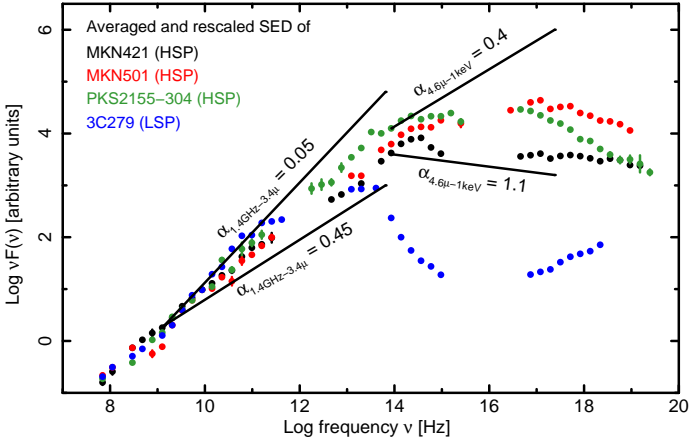}
    \caption{Average and rescaled to 10$^{10}$ Hz SEDs of three well known and representative HSP blazars: MKN 421, MKN 501, and PKS 2155$-$3304, and one LSP, 3C279. The solid lines
    represent the radio to infrared and infrared to X-ray  slope limits used in our selection criteria which are highly effective in differentiating HSP- from LSP-like SEDs. 
    }
      \label{31}
\end{figure}

The parameter ranges given above are derived from the shape of the SED of HSP blazars, which is assumed to be similar to those of the three well known HSPs, i.e. 
MKN 421, MKN 501 and PKS 2155$-$304 shown in Fig. \ref{31}, which also displays the limiting slopes ($\alpha_{1.4GHz-3.4\mu m}$ and $\alpha_{4.6\mu m-1keV}$) used for the selection. 
The condition on $ \alpha_{3.4\mu m-12.0\mu m}$ is used to exclude low energy peaked (LSP) blazars \citep[see ][for details]{massaro_color}. These multi-frequency restrictions drastically reduce the size of the sample from 4.8 million IR sources to only 1347 blazar candidates. These were then studied individually to clean the sample, leaving only HSP sources and HSP candidates.

It is important to stress that one of the main goals of this paper is to test the potential of our selection method, not just to assemble the largest possible sample of HSP blazars. We realise that we are far from being able to select all HSP objects present in the ALLWISE catalogue, mainly due to two reasons: one is that in order to be complete, we would need deep X-ray coverage for the whole sky. Unfortunately, the available catalogues that are deep enough for our purposes (SWIFT, XMM, Chandra) all have very limited sky coverage ($\approx$10-15\%). So, we lose many faint objects with X-ray fluxes below the shallow sensitivity limits of the ROSAT all sky survey. The second main reason is that the IR colour-colour selection demands the sources to be detected in the $3.4\mu m, 4.6\mu m,$ and $12.0\mu m$ WISE channels, a requirement that is less and less fulfilled by faint sources, especially in the WISE longer wavelength channels.

In fact, even with the selection based on the Sedentary IR colours, 46 HSPs from this survey still remain unselected (about 30\% of the sample). Of these, 36 were left out due to bad photometry $˜snr < 2.0˜$ (from which 6 were not detected in any of the WISE channels), and 10 were not selected because they fail to pass the slope criteria requirements. 

\cite{massarotev} report about the selection of a sample of BL Lacs candidate for TeV observations based on the WISE catalogue of infrared sources.
Similarly to our approach these authors searched for blazar candidates in a portion of the colour - colour plane that is wider than the original  WISE blazar strip of  \cite{massaro12} but accept only  X-ray counterparts that 
are strong enough to be listed in the ROSAT Bright Source Catalogue.

\subsection{Deriving the synchrotron peak frequency and classifying the sources}
\label{DerivingPeak}

To make sure that all sources in our sample are HSPs we built the radio to \gr\ SED of each object using the ASDC on-line SED builder tool\footnote{http://tools.asdc.asi.it/SED}, which gives access to 
multi-frequency flux measurements from a large number of catalogues and databases. We determined \nupeak\ by fitting a third degree polynomial function to the data that can be associated to synchrotron emission. Adding the giant elliptical template to the SED (using tools available in the ASDC on-line SED builder tool) we could check whenever the IR and optical data were due to the host galaxy, excluding them from the fit in case of contamination.
When available, XRT and UVOT data (obtained from the SWIFT public archive) were added to the SEDs, to better characterise the $\nu_{\rm peak}$ and $\nu$ f$_{\nu}$ values. 
In almost all cases the available data were sufficient for a good determination of \nupeak. As an example, Fig. \ref{GalaxyContribution} shows the spectral energy distribution of MKN 501 (1WHSP J165352.2+394536), illustrating how \nupeak\ and \nufnupeak\ are determined through a third degree polynomial fit (red solid line) to the data associated to the synchrotron emission. The large variability, a defining feature of blazars, clearly 
plays an important role in the determination of SED parameters. In the specific case of MKN 501, both \nupeak\ and \nufnupeak\ change significantly, illustrating the uncertainties that are intrinsic to these measurements. 
Our polynomial fits are applied to all available data and therefore the parameter estimations reflect the average value of all the flux measurements in the database, smoothing out the effect of variability.

Sources can be classified as HSPs only if they have $\nu_{\rm{peak}} >10^{15}~$Hz. During our detailed visual inspection work, a number of candidates (about 365) were removed from the final sample for a variety of reasons: nearly 27$\%$ of them had not enough data to allow us to estimate \nupeak, about 33$\%$ were removed because they were either identified with known FSRQs (and therefore very likely LSPs) or the fit showed $\nu_{\rm{peak-obs}} < 10^{15}~$Hz, and a few cases (about 6$\%$) were radio extended and therefore likely misaligned jets. Another reason for removing sources was source confusion and the corresponding mismatch of the radio/optical/X-ray positions, especially for faint objects having larger error boxes associated with the radio (SUMSS) and X-ray coordinates. 
After analysing all the 1347 SEDs case by case, the selection criteria listed above gave a clean sub-sample of 850 confirmed HSP blazars or blazar candidates. Therefore, considering that many of the uncertain cases could well turn out to be HSPs, our automatic search method (based on WISE colours and SED slopes) may reach $\approx 63\%$ efficiency.
    
\begin{figure}[h]
   \centering
    \includegraphics[width=1.0\linewidth]{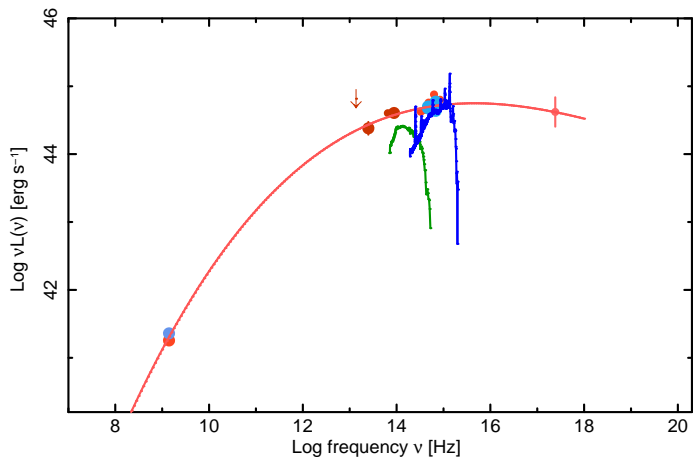}
     \caption{Radio to X-ray SED of 1WHSP153254.8+322250. The green points represent the contribution from the host galaxy assuming a giant elliptical, the blue points are the expectations of a standard QSO accretion (blue bump) rescaled to match the optical points. The red line is an assumption about the non-thermal emission based on the radio, IR and X-ray data, which could be contaminated by the blue bump.}
      \label{seddilutedline}
\end{figure}

\begin{figure}[h]
   \centering
    \includegraphics[width=1.0\linewidth]{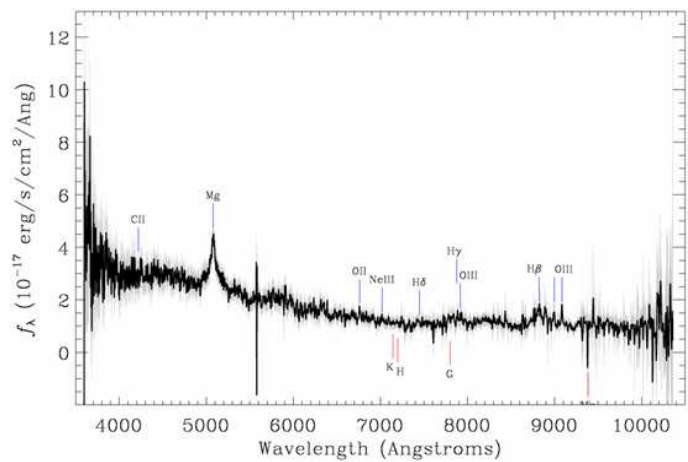}
     \caption{Optical spectra (taken from SDSS-DR12) of 1WHSP153254.8+322250, an object showing broad lines and a possible strong dilution by a non-thermal component in its SED (see Fig. \ref{seddilutedline})}
      \label{dilutedline}
\end{figure}

The above results give us confidence that the majority of our candidates ($\approx$80\%) are indeed genuine HSPs but also tell us that we should expect contamination by spurious sources or LSP FSRQs that appear as HSPs, since the emission from the blue bump sometimes mimics synchrotron emission and 
alters the determination of $\nu_{\rm peak}$, as illustrated in Fig. \ref{seddilutedline}. In fact, considering the latest Sloan Digital Sky Survey\footnote{http://www.sdss.org/} (SDSS) data releases DR10-DR12, 32 previous candidates now have an available spectrum and 7 have been removed from the 1WHSP sample since they presented blue bump features. On average there may be around 21\% contamination of our 416 candidates, translating into $\approx$10\% contamination of the whole 1WHSP sample. 

Finally, we should find more examples of the so far elusive FSRQ/BL Lac transition HSP blazars (Fig. \ref{dilutedline}) where the underlining broad lines are not completely swamped by non-thermal emission from the jet, as predicted by \cite{giommisimplified} and \cite{giommisimplified2}.

\section{The catalogue}\label{catalog}

\subsection{List of HSP blazars and candidates}
\label{totalList}

The procedure described in Sect. \ref{SampleDefinition}, based on the latest version of the WISE catalogue (ALLWISE), led to the selection of  850 sources. However, our first attempt in developing an efficient selection scheme made use of the previous WISE All-Sky Survey catalogue\footnote{In this case, the colour domain explored was delimited by the corners \textit{($c_{4.6-12}$,$c_{3.4-4.6}$)}: (1.42,0.44) (2.05,0.28) (3.81,1.17) (3.29,1.67)}. 
We found that 99 confirmed HSPs that were present in the 1$^{st}$ selection are not present in the 2$^{nd}$, largely because they have W$_{3~snr}<2.0$ in the latest version of the WISE catalogue. These have been added to other known HSP sources that were not selected by our scheme but are included in existing catalogues of blazars. 

To increase the completeness of our sample we added 142 extra HSPs to the 850 sources selected in Sect. \ref{SampleDefinition}. The extra objects are composed by 40\footnote{Of the 46 HSPs that were missed by the selection scheme, 6 lacked detection in all the 
WISE channels} Sedentary sources (SHSP) that were not selected because they do not meet the W$_{3~snr}>2.0$ requirement despite their presence in the WISE catalogue, the 99 confirmed HSPs selected during a first selection as explained above, and 3 confirmed HSP-TeV emitters (see section \ref{TeV}). The final catalogue of HSP blazars and blazar candidates includes 992 sources and is presented in Table \ref{table1}.

The table gives the position taken from the WISE catalogue in column (1), a flag indicating the presence of 
the object in BZCAT in column (2), log($\nu_{\rm{peak-obs}}$) and the logarithm of the 
flux density at the peak (in erg/cm$^{2}$/s) in columns (3) and (4). If the redshift is available, it is 
given in column (5). The gamma-ray counterpart in the {\it Fermi}-
LAT catalogues (1FGL, 2FGL, 3FGL) is given in column 
(6) and in column (7) the photon index $\Gamma$ corresponding to the energy range 0.1GeV$<E<$100GeV is given. Column (8) specifies if the source is a confirmed or a candidate HSP. Within the confirmed 
HSP we distinguish between those that were selected from the SWCD and those added from the Sedentary survey or 
in the TeVCat but not selected by the WISE colours. The HSPs that are flagged as extended sources in the WISE catalogue 
are marked accordingly. Finally, column (9) specifies if the
source is already detected in the TeV band, and column (10) gives a figure of merit (FOM) for a possible
TeV detection (for details on how the FOM is defined, see section \ref{TeV}).

In total, the 1WHSP catalogue contains 992 entries, of which 425 sources are already known blazars (included in the BZCAT catalogue of \cite{5BZcat}, in the Milliquas\footnote{quasars.org/milliquas.htm} list or flagged as confirmed BL Lacs in NED); 
151 are new spectroscopically confirmed HSP blazars based on SDSS or other on-line data, and 416 objects show a blazar-like SED but for which no optical spectrum is available yet. The optical identification of these HSP 
candidates will likely result in a considerable number of sources to be added to blazar catalogues.

\subsection{Lower limits on redshifts} 
\label{zlowerlimits}

In many cases BL Lac spectra are completely featureless. This can be explained by a scenario 
where any emission line or host galaxy light is completely swamped by 
the non-thermal jet component \citep[e.g.][]{giommisimplified}. Following the work of \cite
{landt02} and \cite{SedentaryIII}, we assume a giant elliptical host galaxy to be present in every blazar
that would leave no imprint on the optical band if the total observed flux f$_{obs}$ exceeds the galaxy flux f$_{ell}$ by at least a factor of 10. Therefore, a lack of redshift allows us to set a lower limit  based on the apparent magnitude $m$ in the optical band.

Typically, the mean absolute R-magnitude of a giant elliptical host galaxy within $z<0.6$ is $\langle M_
{R} \rangle_{BL Lacs} =-22.9$ \citep{elliptical}.
The \textit{distance modulus} is given by $(m - M) =  5 log( d_L )  +  25   + k_{\nu , z} $, where 
$d_L$ is the luminosity distance (in Mpc) and $k_{\nu, z}$ is the K-correction, which is written as:

\begin{equation}
\setcounter{equation}{4}
k_{(\nu , z)} = -2.5 log \Bigg( \frac{\nu (1+z) L_{\nu(1+z)} }{\nu L_{\nu}} \Bigg) 
\label{eqM6} 
\end{equation}

The $K$ term depends on the spectral shape of the source L$_{\nu}$; in this case we used the elliptical galaxy template of \cite{EllipticalTemplate}. In particular the R-band was chosen as the reference band, therefore in eq. 4 we have $L(\nu) = L(\nu_{R})$. 
The value of $L_{\nu(1+z)}$ is determined from the host galaxy template and written as a function of z according to $\nu=\nu_R(1+z)$. Since the template is only defined in the range $14.13 < log (\nu) < 14.91$, we can reach $z _{ll} \simeq 0.7$ at maximum.

Considering only the cases where the optical spectra show no features and  
applying the minimum criterion that would generate such featureless profile, that is $f_{obs} > 10 \times f_
{ell}$ we rewrite the distance modulus as eq. \ref{eqM5}. Together with the luminosity distance $d_{L}(z)$ we define a relation to be solved numerically for each $z$, allowing us to calculate the $z_{lower-limit}$ corresponding to the observed $m_R$.

\begin{equation}
\label{eqM5}
m_{R} - \langle M_{R} \rangle_{BL Lacs} = 5 log( d_L (z_{ll})) - 2.5~log(10)  +  25  + k_{(\nu , z)}.
\end{equation}

The $m_{R}$ were retrieved from USNO-B1.0 and from ESO online Digitalised Sky Survey according to availability, making use of the joint interface with the \textit{Data Explorer} tool\footnote{http://tools.asdc.asi.it} that allows one to match radio-optical counterparts. We also
used lower limits to the redshift for the sources included in the \cite{bllacz2}, \cite{bllacz1} and \cite{pita2013} catalogues. In the end, a total of 119 HSPs were assigned a lower limit on their redshift, reaching values as high as $z_{ll} = 0.85$.

To illustrate the method, we consider the case of $m_R$=14.0 and $\langle M_{R} \rangle_{BL Lacs}$=-22.9, for which we derive z$_{ll}$=0.13. Taking into account that the absolute magnitude of elliptical galaxies $M_R$ within $z<0.6$ span the range $-21.5 < M_R < -23.8$ \citep{elliptical}, the redshift lower limits for both extremes (faintest and brightest) are $z_{ll} = 0.077$ and $ z_{ll} = 0.19$, respectively. 
Our calculations, which are based on the assumption that $M_{R}=\langle M_{R} \rangle_{BL Lacs}$, in general produces conservative values. For example, in the case of BZBJ0033-1921 (1WHSPJ003334.3-192132) with m$_{R}$=16.6 the lower limit calculated by our method gives z$_{ll}$=0.32 whereas the spectroscopy method of \citet{pita2013} gives z$_{ll}=0.506$; the same happens for other z$_{ll}$ spectroscopically determined by \cite{bllacz2}.

\subsection{The Selection Efficiency for HSP TeV detected sources}

At the time of writing, the list of sources detected in the TeV band includes 155 objects\footnote{http://tevcat.uchicago.edu} and is rapidly growing thanks to the increasing sensitivity provided by the implementation of stereoscopic arrays of Imaging Atmospheric Cherenkov Telescopes (IACTs) like HESS, VERITAS and MAGIC-II, which are flux limited to $\approx1-3 \times 10^{-13}$~erg/cm$^2$/s at 1 TeV, considering 50h of exposure \citep{Riegertev}. 
About one third (50) of the known TeV sources are located outside the Galactic plane ($\vert b \vert>$20$^{\circ}$), like our sample of HSP blazars. These include 4 low/intermediate synchrotron peaked (LSP/ISP) blazars, 2 pulsar wind nebula, 1 superbubble, 2 starburst galaxies, 2 radio galaxies, and 3 FSRQs.  Of the remaining 36 TeV sources, all classified as HSPs, 33 were included in the 1WHSP sample directly by our selection scheme, which translates into an $\approx 91\%$ selection efficiency of HSP-TeV sources at $\vert b \vert>$20$^{\circ}$. This is a considerable improvement compared to the \textit{Sedentary} survey, which includes only 13 TeV detected sources. 
The 3 TeV detected blazars that our selection criteria missed (BZBJ0303-2407, BZBJ1217+3007 and BZBJ1427+2348) have SEDs with $ \nu_{peak} \sim 10^{15} $ Hz, that is borderline between an ISP and HSP classification, which in these particular cases is strongly affected by flux variability. These three sources were included in the 1WHSP catalogue since their mean SEDs (smoothing the variability along time) are characterised by $ \nu_{peak} \geq 10^{15}$ Hz, and therefore they are likely to be HSPs with an IR counterpart seen by WISE. 

\section{Discussion}\label{discussion}

\subsection{The synchrotron $\nu_{\rm peak}$ distribution}

\begin{figure}[h]
   \centering
    \includegraphics[width=1.0\linewidth,angle=0]{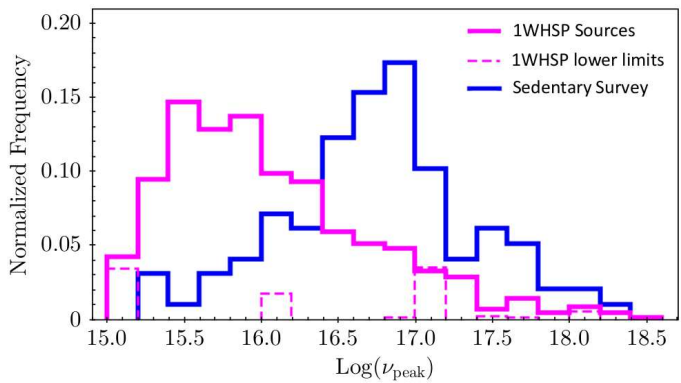}
     \caption{The distribution of the {\it observed} synchrotron peak energy in the full sample of 1WHSP blazars (magenta) and in the Sedentary survey (blue).}
      \label{nupeak_distribution}
\end{figure}

Figure \ref{nupeak_distribution} displays in magenta the distribution of the observed values of \nupeak\  for our 1WHSP sample. The peak of this distribution is near $10^{15.5}~$Hz, reflecting the cut imposed by our selection criteria, and likely some incompleteness near 
the limit of $10^{15.0}~$Hz. 
In the SED of a very small percentage of objects, the X-ray measurements were not sufficient to determine $\nu_{\rm peak}$  because the X-ray flux was still increasing in $\nu$ vs. $\nu f_{\nu}$ space at $\nu \approx  2 \times 10^{18}~$Hz. For these rare cases, and for those for which only \nupeak \ lower limits could be calculated, we plot the distribution as a dashed line.
For comparison, Fig. \ref{nupeak_distribution} shows the $\nu_{\rm peak}$ distribution for the Sedentary BL Lacs (in blue), which instead shows a maximum at $\approx 10^{16.9}$Hz, consistent with the fact that  the Sedentary survey focused on the most extreme HSPs. In both surveys the maximum observed values of \nupeak\ are at $\approx 2-3\times 10^{18}~$Hz. 

There are a few extreme cases where $\nu_{\rm peak-rest}$ (\nupeak\  in the rest frame of the blazar) is larger than $10^{18}$ Hz. One example of a well defined SED (where the X-ray peak is covered by SWIFT's XRT and BAT instruments) is that of WISEJ102212.62+512400.5 at  
z=0.14 with $\nu_{\rm peak-rest}  \sim 2 \times 10^{18}~$Hz. Given the many sources with unknown redshift, $\nu_{\rm peak-rest}$ might reach even larger values.  

The largest observed values of  $\nu_{\rm peak-rest}$ set strong constraints on the maximum energy at which electrons can be accelerated in blazar jets. Assuming a standard SSC model, a magnetic field of B  = 0.1 Gauss, and a Doppler factor = 10, a maximum value of $\nu_{\rm peak-rest}$ of 5 $\times 10^{18}~$Hz  translates into a Lorentz factor of $\sim 10^{6}$.

\subsection{The redshift distribution and the blazar sequence}

Our 1WHSP sample includes 576 {\it confirmed}  blazars, 337 of which ($\sim 58\%$) have reliable redshift measurements, covering the 0.03 -- 0.77 range. 197 ($\sim 34\%$) have instead featureless optical spectra, therefore no redshift determination, and another 42 sources have redshift measurement flagged as {\it uncertain} in the literature. For 114 sources with no redshift it was possible to assign z$_{ll}$ as discussed in sec. \ref{zlowerlimits}, and for another 12 we could only assign upper limits based on  \cite{bllacz2}, \cite{pita2013}, \cite{pks1424p240HighZ}, \cite{PG1553}, \cite{shaw2}, \cite{masetti2013} and \cite{bll}. There still remains 71 objects ($12\%$ of the confirmed HSPs) for which no z nor z$_{ll}$ could be estimated.

Most of the redshift values come from BZCAT \citep{Roma-BzcatI}, from the recent optical spectroscopy work of \cite{bllacz2}, \cite{masetti2013}, \cite{pita2013}, from the SDSS Data Release 12, and from NED. Even though most of the previously identified HSPs showed BL Lac-like spectra, reliable redshift measurements are available when absorption features and/or the Ca H\&K break are visible. In some cases, previous observing campaigns were designed to detect galaxy absorption/emission features \citep[e.g.][]{bllacz1}; Whenever possible, we  incorporated these redshift measurements in our database. 

\begin{figure}[h]
   \centering
    \includegraphics[width=1.0\linewidth,angle=0]{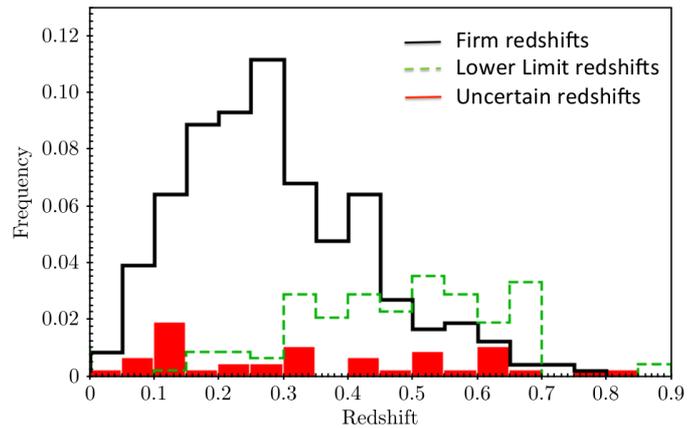}
     \caption{The redshift distribution for the subsample of HSP blazars for which a redshift is known, including uncertain values and lower limits (see text for details).}
      \label{redshift_distribution}
\end{figure}

Figure \ref{redshift_distribution} shows the redshift distribution for our HSP sample, which has $\langle z \rangle =0.28$ and $\sigma=0.14$, 
similar to that reported for BZCAT edition 5.0.0 ($\langle z \rangle \simeq 0.30$) for the subsample of BL Lacs with \nupeak $> 10^{15}$ Hz but lower than the value of  $\langle z \rangle \simeq 0.36$ observed in the subsample of BL Lacs with  \nupeak $< 10^{15}$ Hz in BZCAT5\citep{5BZcat}.
 Figure \ref{redshift_distribution} includes also the distribution of lower limits (see section \ref{zlowerlimits}), which is clearly shifted to 
larger redshift values than the observed ones.  A Kolmogorov-Smirnov test gives a probability of $6\times 10^{-30}$ that the two distributions come from the same parent distribution. This implies, with very high confidence, that the objects with featureless optical spectra must be on average at larger distances compared to the sources with measured redshifts, in full agreement with the predictions of the simplified view of blazars put forward by \cite{giommisimplified, giommisimplified2}.

It is worth recalling that the large \nupeak ~values of HSPs translate into relatively high 
non-thermal jet components at IR and optical wavelengths, which might swamp any accretion disk or host galaxy emission, 
making the redshift determination very difficult or even impossible. Therefore, the redshift distribution from such a population is often incomplete. Said otherwise, high power -- high 
\nupeak~blazars are very hard to identify because they tend to show featureless optical spectra and the consequent lack of redshift hampers any estimation of the emitted power. 
This is because when both radio power and \nupeak~are large, the dilution by the non-thermal continuum becomes extreme and all optical features are washed away \citep[e.g.,][]{giommisimplified,Padovanibllac}. 

\cite{padovanihsp} have shown that the claim of the existence of a blazar sequence, that is of 
a negative correlation between bolometric luminosity and $\nu_{\rm peak}$, might have also
been based on this effect, which could have led to an artificial lack of sources in the high 
power -- high \nupeak~region (see their Fig. 6). Our new sample of HSP blazars provides further strong support 
to this idea. In order to compare 
the bolometric luminosity, $L_{bol} = \nu L_{\nu^S}+\nu L_{\nu^{IC}}$, of our sources with that of 
other blazars, we have taken $L_
{bol}$ to be $1.5 \nu L_{\nu^S}$ \citep{giommisimultaneous}. This correction had to be applied 
since we lack high energy data and cannot determine properly the IC peak for our SEDs. 

\begin{figure}[h]
  \centering
   \hspace*{-0.6cm}\includegraphics[width=1.0\linewidth,angle=0]{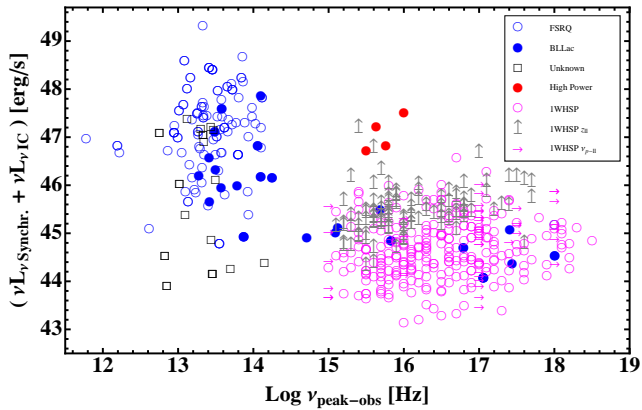}
    \caption{Bolometric luminosity vs. $\nu_{\rm peak}$ for our sample (magenta empty circles) compared with the sample reported by \cite{giommisimultaneous} (blue open and filled symbols) and the high-power HSP from \cite{padovanihsp} (red filled circles). Sources with only lower limits for the redshift are plotted as upward arrows, while sources with lower limits for \nupeak\ are shown as right pointing arrows.}
     \label{blazseq}
\end{figure}

As shown in Fig. \ref{blazseq}, the new HSP sample adds many sources populating the region 
with high $L_{bol}$ and high \nupeak. 
Sources with lower limits on their redshift \citep[calculated or taken from][]{bllacz2} are shown with upward 
arrows. 

Those objects likely populate the upper right quadrant of the plot where high luminosity HSPs 
similar to the four sources discovered by  \cite{padovanihsp} are located.  As examples we 
mention the sources 1WHSPJ142238.8+580155 with z=0.702  \citep{bllac}, PG1553+113 
(1WHSPJ155543.0+111124) with z $>$ 0.433 \citep{PG1553}, and PKS1424+240 
(1WHSPJ142700.4+234800), which is currently one of the TeV detected source with the largest redshift 
limit z $>$ 0.6035 \citep{pks1424p240HighZ}. These high redshift HSP BLLacs, all characterised 
by $L_{bol} \geq 10^{46.5}$ erg/s, are sometimes referred to as \textit{High Power HSPs}. 
Using detailed simulations, \cite{giommisimplified} have predicted the redshift distribution 
of BL Lacs without redshift, which peaks at $z \approx 1.5$ and reaches $z = 3$. Assuming that
one of our most luminous sources (1WHSP J151747.5+652523) were actually at such
redshift, it would have $L_{bol} \sim 10^{48}$, and therefore would be placed in the upper right corner, which would be forbidden in the blazar sequence scenario. This could happen for any of the sources with associated lower limits on redshift, since all of them have confirmed featureless spectra and could be at large distance.

\begin{figure}[h]
   \centering
    \hspace*{-0.8cm}\includegraphics[width=1.0\linewidth]{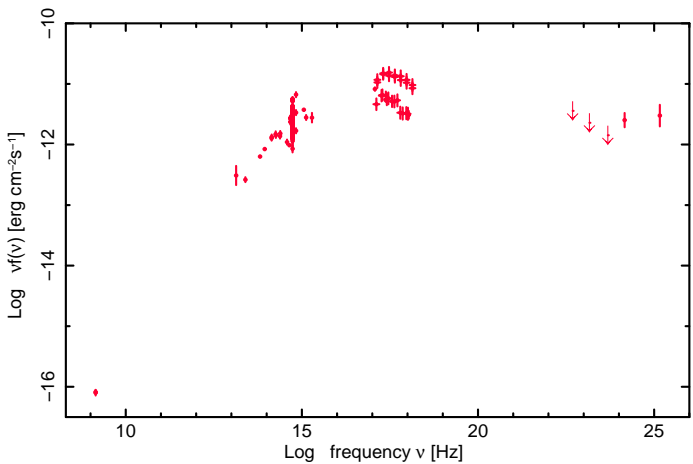}
     \caption{Spectral energy distribution of 1WHSPJ150340.67-154113.0 = 5BZB J1503-1541, an HSP blazar with no known redshift that is detected by Fermi-LAT in the highest energy channels.}
      \label{3}
\end{figure}        

Figure \ref{3} shows a candidate high power HSP (1WHSPJ 150340.6-154113 = 5BZB J1503-1541) which is a confirmed GeV source 
showing hard flux density
up to the highest energy Fermi channels (therefore an extreme object) and has no known redshift, since the spectra is completely featureless. This kind of source could turn out to be a bright and extreme HSP, and probably could be detected in the lower-energy 50-200GeV range by CTA (see sec. \ref{TeVsection}) given that the EBL absorption is not so severe for this energy range. 

\subsection{The infrared LogN-LogS of HSP blazars}
\label{logn-logs}

The 1WHSP blazar sample has been assembled by means of an initial IR colour selection of the high Galactic latitude sources listed in the ALLWISE catalogue, which reduced the sample of potential blazars to about 4.8 million objects. This preliminary selection was 
followed by multi-frequency restrictions and visual inspection to ensure that the SED of all the selected objects is typical of HSP blazars, reducing the sample by more than three orders of magnitude. Additional sources were incorporated, just as described in Sec. \ref{SampleDefinition} and \ref{totalList}, with the restriction that all of them should be IR detected by WISE in the 4.6$\mu m$ channel. Therefore, the final list containing 992 sources is flux limited in the IR band, 
since all the sources must be in the ALLWISE catalogue; however, its level of completeness varies and can be quite poor at low infra-red fluxes where the WISE colours are often not available due to the different sensitivity limits in the three WISE channels.

For the purpose of estimating the IR LogN-LogS we consider the 4.6 $\mu m$ WISE channel (W2) and the subsample of sources that are included in the RASS X-ray survey, which covers the entire sky albeit with sensitivity that strongly depends on ecliptic latitude. 
To estimate the IR LogN-LogS of HSP blazars it is necessary to determine the so-called \textit{sky coverage}, that is the area of sky where the sources could be found for any given IR flux density ($S_{\rm W2}$). 
Since we require that the sources must be included in the RASS X-ray survey, the area useful at any $S_{\rm W2}$ is the  part of the sky where the RASS sensitivity  is better than the flux $f_{limit (1keV)}$, corresponding to  $S_{\rm W2}$, that is  

\begin{equation}
\label{fluxlimit}
\alpha_{4.6\mu m-1keV} = \frac{ log(S_{W2}  /  f_{limit (1keV)}    )    }{   log(   \nu_{1keV} /  \nu_{W2}    )   }  = 1.1  
\end{equation}

see condition 2 in Section \ref{SampleDefinition}

\begin{figure}[h]
   \centering
\includegraphics[width=1.0\linewidth]{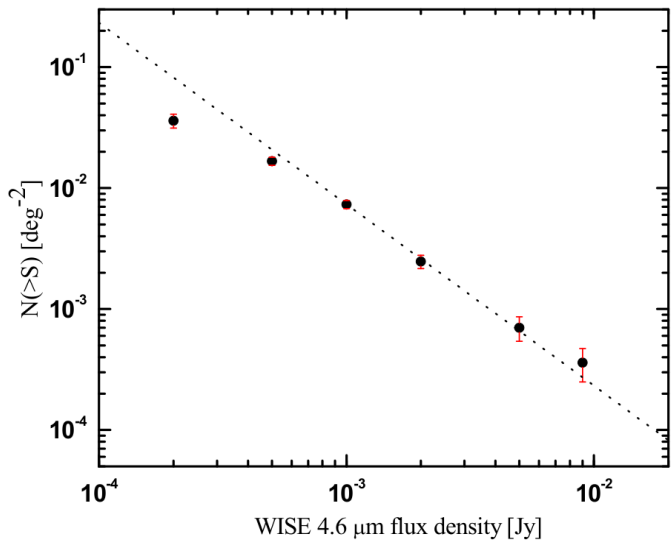}
     \caption{The IR (4.6 $\mu$m) integral LogN-LogS distribution of HSP blazars. The dotted line, corresponding to $N(>S)\propto S^{-1.5}$, represents the expected distribution of a population of sources with constant density in an Euclidean universe. The flattening of the measurements compared to the straight line is, at least partly, due to sample incompleteness near the flux limit.}
        \label{logns}
\end{figure}

In practice, we have chosen six flux density values for $S_{W2}$ (see Table \ref{tablecounts}) and for each of these we calculated the corresponding X-ray flux density limits $f_{limit (1keV)}$. Then we determined the area of the sky covered by RASS (A$_{RASS}$) with sensitivity better than $f_{limit (1keV)}$ and the corresponding number of HSPs per square degree N($>$S$_{\rm W2}$).
The integral LogN-LogS plot (Fig. \ref{logns}) displays a linear trend with slope 1.5, the "Euclidean slope", up to $\lsim 5\times 10^{-3}$ Jy. The LogN-LogS flattens at lower flux densities 
reflecting the onset of severe incompleteness at faint fluxes.

\begin{table}[h]
\caption{ Parameters for calculating the IR LogN - Log S.}       
\centering
\begin{tabular}{c|cccc}
 $S_{\rm W2}$  & $f_{limit (1keV)}$ & A$_{RASS}$ & N$_{sources}$ & N($>$S$_{\rm W2}$)\\ 
 { [mJy]} & [$\mu$Jy] & [deg$^{2}$] &   & 10$^{-3}$[deg$^{-2}$]\\        
\hline
 9.0          & 1.059                      &  27098              &  9        & 0.33 \\   
 5.0          & 0.589                      &  27054              &  21      & 0.77 \\ 
 2.0          & 0.235                      &  26276              &  74      & 2.81 \\ 
 1.0          & 0.117                      &  22796              &  182    & 7.98 \\ 
 0.5          & 0.059                      &  10175              &  221    & 21.7 \\ 
 0.3          & 0.035                      &  4255                &  149    & 30.5 \\
 0.2          & 0.023                      &  1578                &  76      & 48.1
\end{tabular}
\label{tablecounts}
\end{table}

 If we assume that, after correcting for incompleteness, the Euclidean slope can be extrapolated down to S$_{4.6 
\mu m} \sim 10^{-4}$~Jy,  the expected density of HSP blazars is $\approx 0.2$ deg$^{-2}$, 
corresponding to a total of $\approx$ 8,000 objects in the sky that may be within the detection capabilities of 
next generation of VHE detectors. The IR LogN-LogS for HSP blazars, combined with the distribution of \nupeak\ shown in Fig. \ref{nupeak_distribution}, can be used to estimate the contribution of such sources to the cosmic X-ray Background (CXB) light, especially because the integrated flux of blazars may produce a considerable fraction of it, as shown by \cite{giommicmb}. Moreover, HSPs 
are the dominant extragalactic sources of TeV photons, and a LogN-LogS may provide a way to estimate the intensity of a putative TeV background. 

\subsection{The $\gamma$-ray spectral index vs $\nu_{\rm peak}$ }

\cite{abdo10} showed that the \gr\ spectral index of blazars is correlated with \nupeak, which
they estimated using an approximate method based on radio, optical and X-ray flux density ratios.

\begin{figure}[h]
   \centering
    \hspace*{-0.4cm}\includegraphics[width=1.0\linewidth]{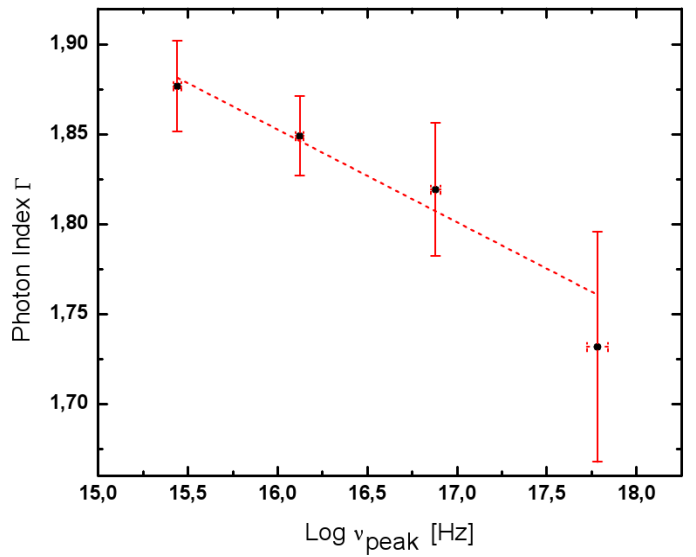}
     \caption{The average Fermi-LAT $\gamma$-ray photon spectral index of all HSPs in our sample detected by Fermi-LAT, binned in four intervals of log(\nupeak), is plotted as a function of log(\nupeak). The best fit line corresponds to $\Gamma_{Fermi-LAT}$ = -0.051 Log(\nupeak) + 2.67 }
      \label{fermislopes}
\end{figure}

To test this dependence using our sample of HSP blazars we plot in Fig. \ref{fermislopes} the 
average values of the \gr\ spectral index in four bins of \nupeak\ as a function of \nupeak. A clear correlation is present confirming the finding of \cite{abdo10}, and that there is a hardening of the SED slope between 100MeV-100GeV with increasing $\nu_{peak}$ values. Therefore, the most extreme HSPs might be associated with brighter (in situ) TeV sources. In fact, redshift becomes one of the key parameters to be taken into account, since TeV absorption due to interaction with the Extragalactic Background Light (EBL) can significantly shape the SED in the VHE band. An estimate of the TeV-flux based on parameters derived from the SED (like: $\nu^{SC}_{peak}$, $\nu^{IC}_{peak}$, and its respective $\nu f_{\nu}$ values) is not simple and could be explored in future work. 

\subsection{The TeV band}\label{TeV}
\label{TeVsection}

\begin{figure}[h]
 \centering
  \includegraphics[width=1.0\linewidth]{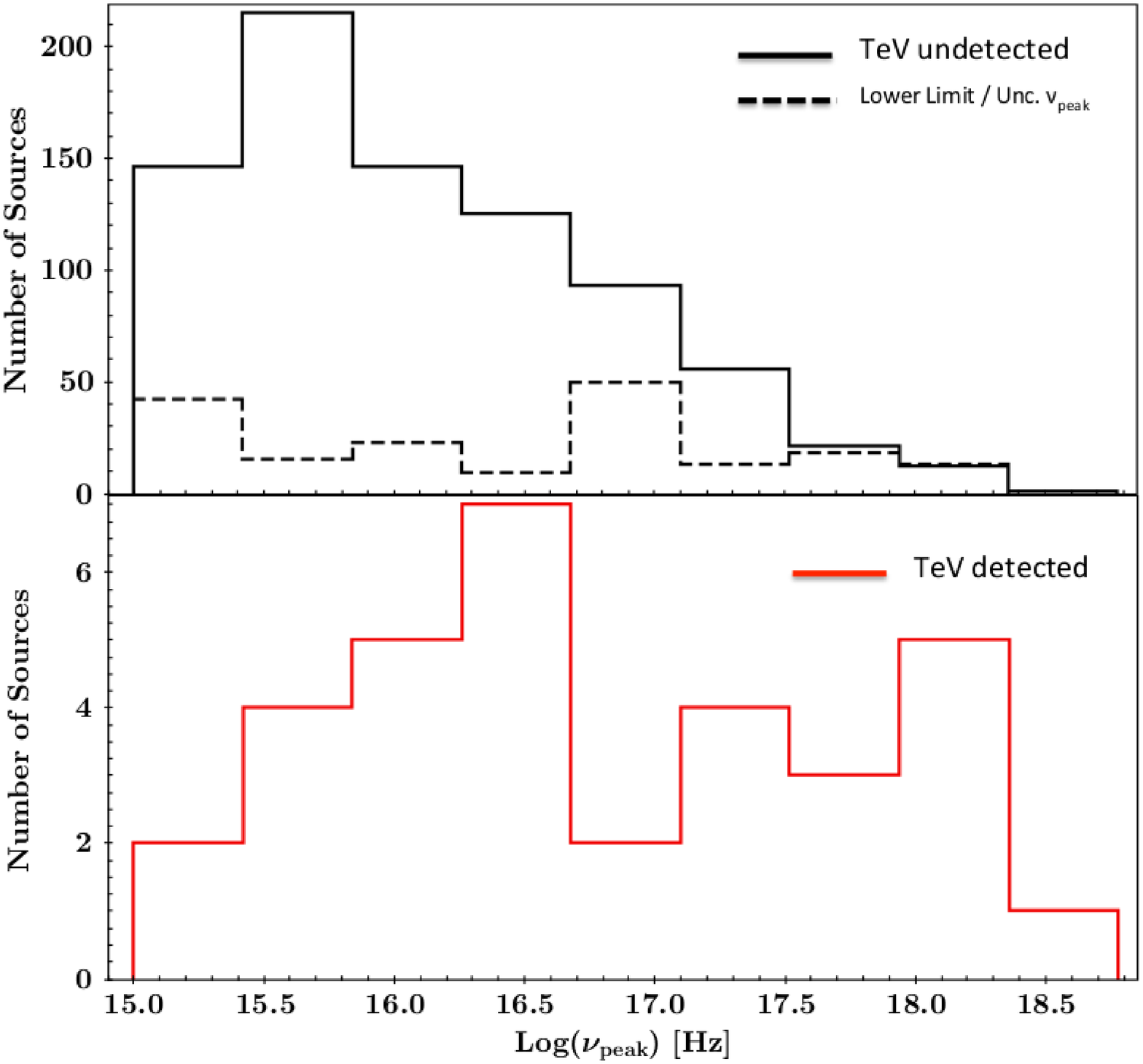}
   \caption{The distribution of synchrotron peak frequencies for the 1WHSP sources that are still undetected in the TeV band (top panel) 
 and those already detected (bottom panel) by Cherenkov telescopes.}
    \label{tevdets}
\end{figure}

Figure \ref{tevdets} shows that the \nupeak\ distribution of the 1WHSP subsample of TeV detected sources (bottom panel) spans the entire range ($ 10^{15} -  \gsim 10^{18}$ Hz) covered by still undetected 1WHSPs (top panel). This implies that all objects in our list could be detected at TeV energies, given that enough sensitivity is achieved. In addition, we note that a few ISP blazars ($10^{14}$ Hz $<$ \nupeak $< 10^{15}$Hz) have been detected by Cherenkov telescopes, extending the above conclusion to lower synchrotron peak energies.

\begin{figure}[h]
 \centering
  \includegraphics[width=0.95\linewidth]{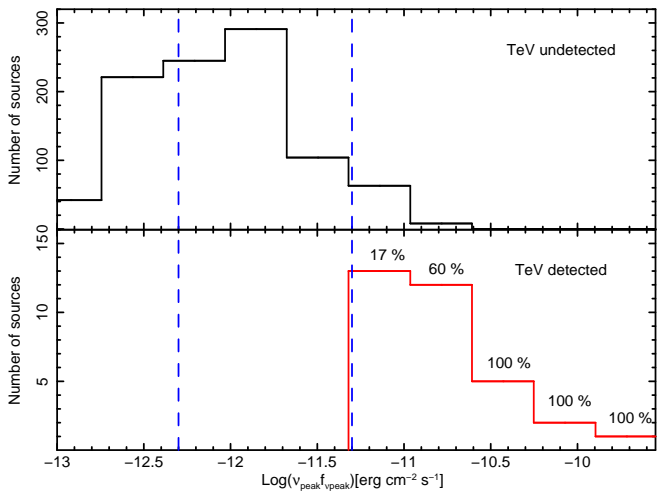}
   \caption{The distribution of synchrotron peak fluxes for the 1WHSP sources that have been detected so far (bottom panel) 
   and that are still undetected in the TeV band (top panel). For each bin in \nufnupeak~we report the percentage of 1WHSP sources that are already TeV detected. Central and left dashed lines correspond respectively to the detectability limit of present Cherenkov Telescopes and approximately to the future CTA.}
    \label{tevflux}
\end{figure}

Figure \ref{tevflux} gives the distribution of the synchrotron peak fluxes, \nufnupeak, of the 1WHSP blazars that have been detected in the TeV band (bottom panel) and of those that are still undetected (top panel). For each bin in \nufnupeak , we have calculated the percentage of 1WHSP sources that are already TeV detected. From this it is clear that the \textit{TeV detected sources} so far are the brightest objects. Note that the peak flux of the undetected blazars in many cases is just below that of the detected ones, and is never more than about a factor ten fainter than the faintest detected object. In Fig.  \ref{tevflux} the central dashed line at $Log(\nu_{peak}f_{\nu peak})=-11.3$ corresponds to the detectability limit of present IACTs. Since CTA will reach one order of magnitude lower sensitivity, it may be possible to detect HSP sources one order of magnitude fainter down to $ Log(\nu_{peak}f_{\nu peak})=-12.3$, corresponding to the left dashed line.

\begin{figure}[h]
 \centering
  \hspace*{0.0cm}\includegraphics[width=1.0\linewidth]{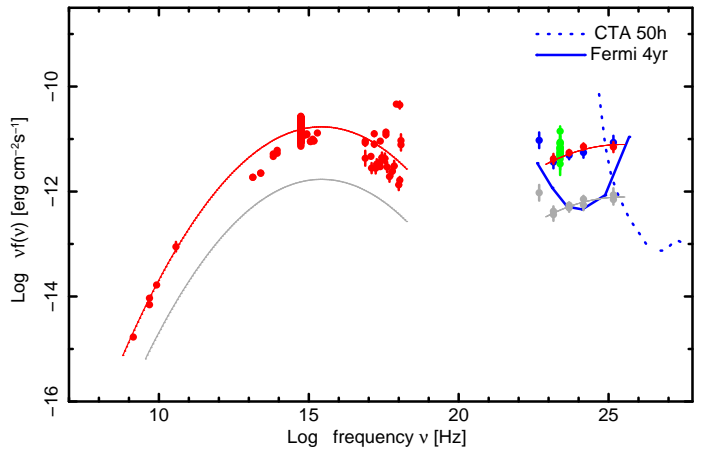}
   \caption{SED 1WHSPJ172504.3+115215. Sensitivities from Fermi-LAT four year exposure represented by the blue line and CTA with 50h exposure represented by the blue dashed line, which are equivalent to each other at $\approx$50 GeV.}
    \label{CTA50h}
\end{figure}

Given that variability of one order of magnitude or even larger is often observed in the X-ray and TeV bands, most of the HSP blazars in our sample (with the exception of those at very high redshift) may be detectable during flaring episodes by the present generation of Cherenkov telescopes, and certainly by the future CTA, as shown in Fig. \ref{CTA50h}. There we present the SED of 1WHSP J172504.3+115215 and compare the sensitivities of Fermi-LAT (for a four year exposure) and CTA (for a 50 hour exposure  \citep{CTA50h}), which are approximately equivalent at the energy of 50 GeV.

Scaling down the SED (red solid line in Fig. \ref{CTA50h}) by one order of magnitude we obtain the grey line, so that the $\gamma$-ray flux approaches the Fermi-LAT sensitivity limit. Despite that, such a faint HSP may be seen by CTA in the lower-energy channels since the mean $\gamma$-ray spectral index for the 1WHSP sources is given by $\langle \Gamma \rangle=1.85 \pm 0.01$ . Therefore, on average, HSPs are characterised by hard $\gamma$-ray spectral index, favouring detectability by CTA in the 0.5-1TeV energy range.

The lower energy threshold of Cherenkov telescopes is decreasing significantly, enlarging their reach to larger redshifts, so we conclude that probably most of our HSPs are good targets for near future VHE observatories. 

To provide a quantitative measure of potential detectability by TeV instruments we introduce a Figure Of Merit (FOM), defined as the ratio 
between the synchrotron peak flux \nufnupeak\ of a given source and that of the faintest blazar in our 
sample that has already been detected in the TeV band. This FOM is reported in Table \ref{table1} for all sources of our 
sample and gives an objective way of assessing the likelihood that a given HSP may be detectable as a TeV source. A total of 112 sources have FOM$\geq$1.0, meaning that their synchrotron peak flux are as bright as the faintest HSP already detected as TeV sources. Note that within the entire 1WHSP sample there are 36 TeV detected sources up to now, which leaves 76 (high FOM) potential TeV sources that may be detectable by the present generation of detectors. This is consistent with the results of \cite{PadGiom2015}, who, by means of detailed simulations, have predicted that $\ge 100$ new blazars can be detected now by current IACTs.

 \subsection{1WHSP objects as possible neutrino sources}

Recently \cite{Pad_Res_2014} have suggested a possible
association between HBL BL Lacs and seven neutrino events reported by the
IceCube collaboration \citep{ICECube14} based on joint positional and
energetic diagnostics. Namely, \cite{Pad_Res_2014} looked for sources
in available large area high-energy $\gamma$-ray catalogues within the
error circles of the IceCube events and then compared the SEDs 
of these sources with the energy and flux of the
corresponding neutrino. We stress that {\it all} HBL BL Lacs in their
list of most probable counterparts (their Tab. 4) are 1WHSP sources
with FOM $\ge 1.2$. This vindicates the use of a selection on
synchrotron peak and flux to identify present or potential high-energy (TeV
or even larger) emitters and opens up the possibility that our sources might 
bridge the gap between ``classical'' and neutrino astronomy.

\section{Conclusions} 

Using the ALLWISE infrared catalogue of sources combined to multi-frequency data we have assembled the largest sample of HSP blazars. This was done for three main reasons:
\begin{itemize}
\item estimate the surface density of HSP blazars down to relatively faint infrared fluxes;
\item study the multi-frequency properties of HSP blazars;
\item build a large catalogue of potential targets for the present and future generations of Cherenkov telescopes.
\end{itemize}

\begin{figure}[h]
 \centering
  \includegraphics[width=1.0\linewidth]{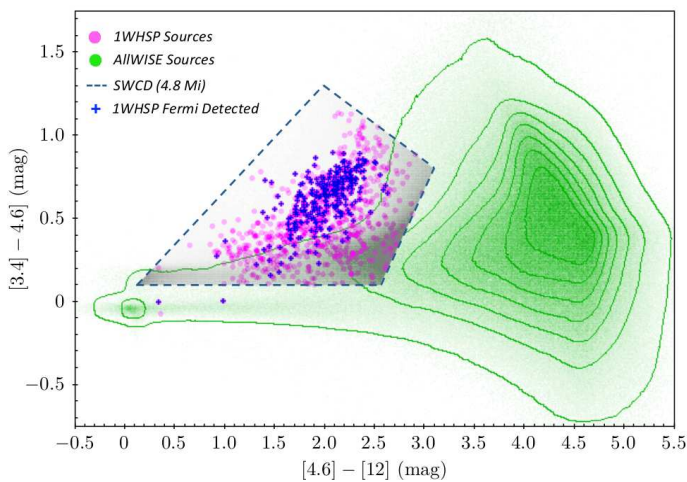}
   \caption{The ALLWISE colour-colour diagram, with 1WHSP sources (magenta) highlighting the ones having a $\gamma$-ray counterpart (blue cross), and the SWCD delimited by the dashed lines.}
    \label{1WHSPcolours}
\end{figure}

The initial selection of the sample (similarly to \citet{massarotev}) was done in the WISE (W2-W3 vs W1-W2) colour-colour space  taking sources in the area that encompass all the Sedentary HSPs included in the ALLWISE catalogue and detected with snr $\geq$2.0. 
Placing all the 1WHSP sources of our sample in the colour-colour diagram (Fig. \ref{1WHSPcolours}) gives an overview of their distribution within the SWCD, highlighting the cases having confirmed $\gamma$-ray counterparts in 1/2/3FGL catalogues.

Some examples of the so far elusive FSRQ/BL Lac transition HSP blazars (where the broad lines are not completely swamped by non-thermal emission from the jet) predicted by \cite{giommisimplified} and \cite{giommisimplified2} may have been found.  

The sample of HSP blazars and blazar candidates presented here contains 992 objects but it is not statistically complete, in the sense that 
not all sources above the WISE flux limit are included. Our IR LogN-LogS results give a surface density of $\sim$ 0.2 HSP blazars 
per square degree near the WISE 4.6$\mu$m flux limit, implying that the total number of HSP sources in the entire sky above the WISE flux limit should be of the order of a few thousands.

In table \ref{table1} we highlight 76 promising TeV candidates with $FOM \geq 1.0$ not yet TeV 
detected. Those should be within reach of present-generation detectors since their synchrotron peak fluxes are as large as those of the faintest TeV-HSP already detected. Moreover, some of 1WHSP sources with $FOM>1.2$ were reported \citep{Pad_Res_2014} as possible extragalactic counterparts for the very high energy neutrinos observed by IceCube.

We called our catalogue 1WHSP, where W stands for WISE, HSP stands for High Synchrotron Peaked blazars, and the  prefix 1 is used since we intend to release updated lists in the future. These will be labelled with progressive numbers and  
will benefit from additional X-ray and optical data as it becomes available, and will include many new objects, mostly because we will base the selection only on radio-infrared-X-ray flux ratios relaxing the requirement that the candidates must be detected in all three WISE bands. 

\begin{acknowledgements}
        
BA and BF are supported by the Erasmus Mundus Joint Doctorate Program by Grant Number 
2011-1640 and 2010-1816 respectively, from  the EACEA of the European Commission. PP thanks the ASI Science Data Center (ASDC) for the hospitality and partial financial support for his visits.  This work was supported by the ASDC, Agenzia Spaziale Italiana; University La Sapienza of Rome, Department of Physics. This publication makes use of data products from the Wide-field Infrared Survey Explorer, which is a joint project of the University of California, Los Angeles, and the Jet Propulsion Laboratory/California Institute of Technology, funded by the National Aeronautics and Space Administration. We also make use of archival data and bibliographic information obtained from the NASA/IPAC Extragalactic Database (NED), data and software facilities from the ASDC managed by the Italian Space Agency (ASI), and TOPCAT software \citep{topcat}. We thank F. Massaro for useful discussions and the anonymous referee for her/his suggestions. The \textit{Fermi} LAT Collaboration acknowledges generous ongoing support
from a number of agencies and institutes that have supported both the
development and the operation of the LAT as well as scientific data analysis.
These include the National Aeronautics and Space Administration and the
Department of Energy in the United States, the Commissariat \`a l'Energie Atomique
and the Centre National de la Recherche Scientifique / Institut National de Physique
Nucl\'eaire et de Physique des Particules in France, the Agenzia Spaziale Italiana
and the Istituto Nazionale di Fisica Nucleare in Italy, the Ministry of Education,
Culture, Sports, Science and Technology (MEXT), High Energy Accelerator Research
Organization (KEK) and Japan Aerospace Exploration Agency (JAXA) in Japan, and
the K.~A.~Wallenberg Foundation, the Swedish Research Council and the
Swedish National Space Board in Sweden. Additional support for science analysis during the operations phase is gratefully acknowledged from the Istituto Nazionale di Astrofisica in Italy and the Centre National d'\'Etudes Spatiales in France.

\end{acknowledgements}

\bibliographystyle{aa}
\bibliography{wisepaper}


\begin{longtable}{p{2.6cm}p{0.3cm}p{1.2cm}p{1.6cm}p{1.9cm}p{3.2cm}p{1.3cm}p{0.6cm}p{0.4cm}p{0.8cm}}
\caption{\label{table1} \textbf{1WHSP sample.} The column \textit{Type} indicates if the source is a spectroscopically confirmed HSP (0 for previously known blazar, $0^+$ for newly identified blazar), a candidate HSP (1), a complementary source from the Sedentary survey (2), if it is classified as extended by WISE (3), and complementary source from TeVCat (4). The column \textit{Fermi-LAT} shows the $\gamma$-ray counterpart in the FERMI catalogs and their respective photon index $\Gamma$ measured for the energy band 0.1GeV$<E<$100GeV. The columns \textit{Bz} and \textit{TeV} show if the source is part of those catalogs. The column \textit{FOM} represents the likelihood of detection in the TeV band (see section 4.5). The lower limit on $\log(\nu_{peak})$ is due to lack of data in higher energies. Upper/Lower limits on the redshift are taken from \cite{bllacz2} (a), \cite{pita2013} (b), \cite{pks1424p240HighZ} (c), \cite{PG1553} (d), \cite{shaw2} (e), \cite{masetti2013} (f), \cite{bll} (g), or calculated by us (see Section \ref{zlowerlimits}) using the observed m$_r$ that is given in lower case. Entries marked with an asterisk are uncertain. \textbf{The full catalog is available at www.asdc.asi.it/1whsp containing complementary information and a direct link to the \textit{SED building tool} where the multifrequency spectra for each source can be quickly visualised.} }\\
\hline       
1WHSPJ & Bz & Log($\nu_{p}$) & Log($\nu f_{(\nu_p)}$) & z & Fermi-LAT & $\Gamma$ & Type & TeV & FOM  \\      
\hline
\endfirsthead
\caption{continued.}\\
\hline\hline
1WHSP J & Bz & Log($\nu_{p}$) & Log($\nu f_{(\nu_p)}$) & z & Fermi-LAT & $\Gamma$ & Type & TeV & FOM   \\     
\hline
\endhead
\hline
\endfoot

    000116.3+293534  &   -           &    $>$17            &     $>$-12.8          &     0              &    -                   &  -               &    1         &   -  &    0.03    \\   
    000132.7-415525  &   -           &      15.8           &       -11.7           &     0              &     2FGL J0001.7-4159  &  2.1$\pm$0.19    &    1         &   -  &    0.40    \\   
    000213.7-103816  &   -           &      16.6           &       -13.1           &     0              &    -                   &  -               &    1         &   -  &    0.02    \\   
    000319.5-524727  &   -           &      15.4$^{*}$     &       -12.4$^{*}$     &     0              &     3FGL J0003.2-5246  &  1.9$\pm$0.17    &    1         &   -  &    0.08    \\   
    000513.7-261438  &   -           &      15.4           &       -12.6           &     0              &    -                   &  -               &    1         &   -  &    0.05    \\   
    000835.3-233927  &   y           &    $>$17            &     $>$-11.8          &     0.147          &     2FGL J0008.7-2344  &  1.6$\pm$0.25    &    3         &   -  &    0.31    \\   
    000949.7-431650  &   -           &      15.6           &       -12             &    $>$ 0.56$_{(18.6)}$        &    -                   &  -    &    0$^{+}$   &   -  &    0.20    \\    
    001011.7+334851  &   -           &      15.2           &       -12.1           &     0              &    -                   &  -               &    1         &   -  &    0.16    \\
\end{longtable}

\label{table}

\end{document}